\documentstyle[aps,prb,multicol]{revtex}
\begin{document}
%\tightenlines
\include{psfig}

\title{\bf Microwave fluctuational conductivity in 
YBa$_{2}$Cu$_{3}$O$_{7-\delta}$}

\author{E.Silva$^{(1)}$, R.Marcon$^{(1)}$, S. Sarti$^{(2)}$, R. Fastampa$^{(2)}$, M. 
Giura$^{(2)}$, M. Boffa$^{(3)}$, A. M. Cucolo$^{(3)}$}

\address{ $^{(1)}$Dipartimento di Fisica ``E.Amaldi'' and Unit\`{a}  INFM,\\
Universit\`{a} ``Roma Tre'', Via della Vasca Navale 84, 00146 Roma, 
Italy\\
$^{(2)}$Dipartimento di Fisica and Unit\`{a}  INFM,\\ Universit\`{a} ``La Sapienza'',
P.le Aldo Moro 2, 00185 Roma, Italy\\
$^{(3)}$Dipartimento di Fisica, Universit\`{a} di Salerno, Baronissi, 
Salerno, Italy}

\draft\date{Dec. 23, 2003; revised version of 
cond-mat/0208343, to be published in 
European Physical Journal B} \maketitle

\begin{abstract}
We present measurements of the microwave complex conductivity at 23.9 
and 48.2 GHz in YBa$_{2}$Cu$_{3}$O$_{7-\delta}$ films, in the 
fluctuational region above $T_{c}$.  With increasing temperature, the 
fluctuational excess conductivity drops much faster than the 
well-known calculations within the time-dependent Ginzburg-Landau 
theory [H. Schmidt {\em Z. Phys.} {\bf 216}, 336].  Approaching the 
transition temperature, slowing down of the fluctuations is observed.  
We discuss the results in terms of a modified gaussian theory for 
finite-frequency fluctuational conductivity, where renormalization is 
introduced in order to account for the $T\rightarrow T_{c}$ regime, 
and a spectral cutoff is inserted in order to discard high-momentum 
modes.  The data are in excellent agreement with the modified theory, 
when formulated for three-dimensional, anisotropic superconductors, in 
the whole experimentally accessible temperature range, and for both 
frequencies.
\end{abstract}

\section{Introduction}
\label{intro}

High-$T_{c}$ superconductors are particularly suitable for the study 
of fluctuational effects: high critical temperatures, short coherence 
lengths and strong anisotropy have as a direct consequence a strong 
widening of the temperature window where fluctuations dictate the 
behavior of various physical observables.  In particular, it is widely 
recognized that the temperature region where critical fluctuations 
might be observable can be of order $\sim$1 K.\cite{lobb} On the other 
hand, the widening of the fluctuational temperature window extends to 
the high temperature region, thus giving the opportunity to test the 
validity range of fluctuational models beyond the $T\rightarrow T_{c}$ 
limit.  It appears that both the regions close to and far from $T_{c}$ 
are worth of being investigated in cuprates.\\
Like in conventional, low-$T_{c}$ superconductors,\cite{skoc} most of 
the experimental data have been collected by measuring dc 
conductivity.\cite{freitas,ausloos,hopf,balest92,labdi,calzona,pavuna,cimberle,silvaPhC,carballeira,silva} 
Finite-frequency experiments, however, are a potential source for 
additional information, such as the estimate of the Ginzburg-Landau 
(GL) relaxation time.  Moreover, a finite-frequency study in the 
fluctuational regime is in principle a more stringent test for 
theoretical models, since at a single frequency two curves (real and 
imaginary parts) have to be fitted by the model with the same 
parameters, and similarly the model must fit to the frequency 
dependence.  A drawback is that measurements at frequencies high 
enough to yield data significantly different from the dc case 
(typically, in the microwave range) are rather difficult to perform 
near the normal state (most of the microwave experiments are optimized 
to small values of the impedance, i.e.  well below the transition).  
In fact, in low-$T_{c}$ superconductors only a very few measurements 
of the complex excess conductivity have been performed,\cite{oldmicro} 
most of them at microwave frequencies or above.  In cuprates, some 
measurements focussed to the temperature region very close to $T_{c}$, 
in order to assess the relevance of critical 
fluctuations,\cite{anlage,booth,neri,waldram} and only a few 
reports\cite{nerisatt,nerimos} explored a wider temperature range 
above $T_{c}$.  In this latter case, a significant reduction of the 
fluctuational conductivity below the theoretical expectations as 
calculated within the time-dependent Ginzburg-Landau theory 
\cite{schmidt} was found.  This feature was connected to the 
progressively smaller contribution of the high-momentum modes to the 
excess conductivity, as the temperature is raised above $T_{c}$: the 
so-called short-wavelength cutoff (SWLC) regime.  Such a physical 
phenomenon has been successfully invoked to describe the excess dc 
conductivity in conventional superconductors\cite{johnson} and 
cuprates.\cite{freitas,hopf,silvaPhC,carballeira,silva} It is then 
interesting to check this approach against a more stringent test, such 
as multifrequency measurements of the complex conductivity.\\
We notice that all the estimates of the excess conductivity (in 
zero magnetic field) rely on the assumption of some kind of normal 
state, existing at temperatures well above $T_{c}$, from which 
superconductivity emerges.  In this paper we will follow the same 
approach, assuming in particular a linear $T$-dependent normal state 
above $T_{c}$, but we mention that alternative scenarios, connected to 
the so-called pseudogap phase, could affect some of the conclusions 
that have been given in all the papers devoted to the zero-field 
paraconductivity.  To mention a few of the possible alternative 
scenarios, intrinsically inhomogeneous 
superconductivity\cite{ovchinnikov} would lead to a dissipationless 
state at the percolation threshold, with excess diamagnetism and 
anomalous frequency dependence of the surface impedance in the 
pseudogap region; preformed pairs well above $T_{c}$, invoked for the 
interpretation of the optical conductivity,\cite{mihailovic} and 
competing order parameters,\cite{chakravarty} with a hidden phase 
transition at the pseudogap $T^{\star}$, could as well increase the 
in-plane conductivity, thus giving an estimate of the paraconductivity 
lower than the one obtained with a linearly extrapolated normal state.  
It is then in any case necessary to consider models that predict a 
reduced paraconductivity with respect to the established theory.\cite{schmidt}\\
In this paper we present measurements of the fluctuational complex 
conductivity $\Delta\sigma$ in two high quality 
YBa$_{2}$Cu$_{3}$O$_{7-\delta}$ (YBCO) thin films as a function of the 
temperature, and for two microwave frequencies (23.9 and 48.2 GHz).  
We show that the well-established theoretical result\cite{schmidt} does 
not describe either the frequency dependent excess conductivity 
approaching $T_{c}$, or the fast drop of the excess conductivity as 
the temperature is raised far from $T_{c}$.  We extend the established 
model \cite{schmidt} to include the short-wavelength 
cutoff,\cite{silvaEPJB} which accounts for the reduced 
paraconductivity far from $T_{c}$, and the 
renormalization,\cite{dorsey,neri} effective close to $T_{c}$.  The 
extended model describes very well the temperature and frequency 
dependence of the complex conductivity.  Several superconducting 
parameters are also evaluated.

\section{Experimental section}
\label{exp}

\begin{figure}
\centerline{\psfig{figure=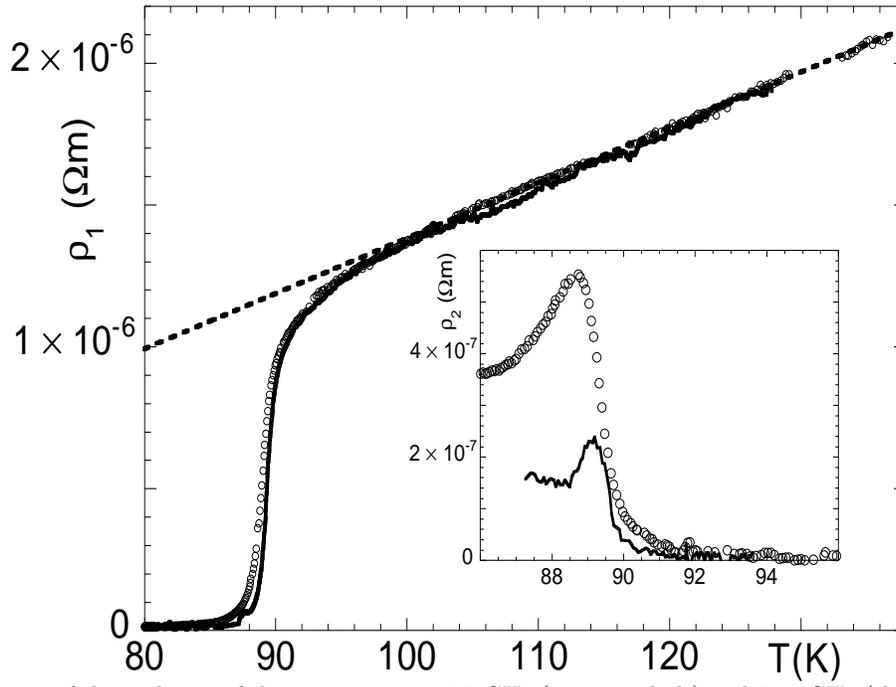,height=9cm,width=12cm,clip=,angle=0.}}
\caption{Measurements of the real part of the resistivity 
at 48.2 GHz (open symbols) and  23.9 GHz (thick line).  
Dashed line: extrapolated normal-state resistivity.  Inset: imaginary 
part of the resistivity.  Symbols as in main panel.  Measurements 
taken on sample YA.}
\label{fig1}
\end{figure}

The complex conductivity was obtained from measurements of the 
temperature dependence of the quality factor and frequency shift of 
cylindrical metal cavities, resonant in the TE$_{011}$ mode.  The 
sample was mounted as an end wall, so that microwave currents flow in 
the (a,b) planes, and measurements yielded the (a,b) plane response.  
The experimental systems, extensively described 
elsewhere,\cite{fastampaMST,silvaMST,silvareview} were modifed in 
order to measure the frequency shift.\cite{nerithesis} Two cavities 
were used, for measurements at $\omega/2\pi$= 23.9 and 48.2 GHz.  Accurate 
calibration of the resonators was performed, in order to obtain the 
effective surface resistance $R^{eff}$ and the temperature variation of the effective 
surface reactance $\Delta X^{eff}$. \\
Two YBa$_{2}$Cu$_{3}$O$_{7-\delta}$ films, with transition 
temperatures $T_{c}\sim$ 90 K, were grown on 1 cm $\times$ 1 cm 
$\times$ 1 mm LaAlO$_{3}$ substrates by planar high oxygen pressure dc 
sputtering technique.  The thickness of the films was $d$=(2200$\pm$ 
200)$\;$\AA. Details of the film growth\cite{beneduce} and 
characterization\cite{beneduce,neriPhC} have been reported elsewhere.  
Typical characteristics of these films are\cite{beneduce} $\Delta 
T_{c}<$0.5 K (from inductive measurements), surface resistance below 
80 $\mu\Omega$ at 77 K and 5.4 GHz, and FWHM of the rocking curves of 
the (005) peak of 0.1$^{o}$ (indicating excellent c-axis orientation).  
X-ray $\Phi$-scan analysis\cite{neriPhC} indicated strong biepitaxiality along (a-b) 
plane.  Average roughness of 20\AA $\;$ over 1 $\mu$m $\times$ 1 
$\mu$m area was determined by AFM. The thickness of the films was of 
the order of the zero-temperature London penetration depth, so that 
close and above $T_{c}$ the thin-film approximation is fully 
justified.  Thus, the measured effective impedance directly yields the 
complex resistivity\cite{silvaSUSTthin} through: 
$R^{eff}\simeq\rho_{1}/d$, $\Delta X^{eff}\simeq\Delta\rho_{2}/d$.  By 
accurate comparison with the cavity background, we did not resolve a 
frequency shift due to the sample above $\sim$95 K. We then assumed 
$\rho_{2}=0$ in our near-optimal doping samples in the normal state 
(recent microwave measurements\cite{kusko} have shown that this {\em 
ansatz} can be violated in underdoped samples, but it is valid in 
optimally doped cuprates), so that from the measured quality factor 
and frequency shift we get the complex resistivity, 
$\rho(T)=\rho_{1}(T,\omega)+{\rm i}\rho_{2}(T,\omega)$.  The 
as-extracted complex resistivity of sample YA is reported in 
Fig.\ref{fig1} for the two frequencies investigated.\\
The complex excess conductivity 
$\Delta\sigma(T,\omega)=\Delta\sigma_{1}(T,\omega)+{\rm 
i}\Delta\sigma_{2}(T,\omega)$was calculated by subtracting the 
temperature dependent normal state real resistivity 
$\rho_{n}(T)=1/\sigma_{n}(T)$, linearly extrapolated above 130 K, 
accordingly to the expression:
\begin{equation}
\rho(T,\omega)=\rho_{1}(T,\omega)+{\rm i}\rho_{2}(T,\omega)=
\frac{1}{\sigma_{n}(T)+\Delta\sigma(T,\omega)}
\label{rosigma}
\end{equation}
We carefully checked that by changing the temperature range of the 
linear fit of the normal state resistivity in the region 125 K -165 K 
we did not affect the obtained $\Delta\sigma$ below 110 K.\\
In doing 
all the necessary transformations from measured quality factor and 
frequency shift to complex excess conductivity, an easily calculated 
geometrical factor is needed.  Errors in the geometrical factor, as 
well as in the estimate of $d$, all reflect on a simple scale factor 
in the calculated conductivity, without affecting the temperature 
dependence.\\
Before discussing the data for $\Delta\sigma$, we summarize the 
theoretical framework that we adopt.\\

\section{Theoretical background}
\label{th}
In this Section we summarize and further extend well known and recent 
results for the complex excess conductivity of an anisotropic, three 
dimensional superconductor.  Explicit calculations, in 3D, 2D and 1D, 
have been reported previously.\cite{silvaEPJB}\\
On theoretical grounds, fluctuational phenomena in superconductors 
have been addressed by microscopic as well as by phenomenological 
theories (see, e.g., the reviews in Ref.s 
\onlinecite{skoc,varlareview,mishonov}).  The time-dependent 
Ginzburg-Landau (TDGL) approach is very useful since it allows for a 
first determination of important parameters, such as the coherence 
length and the relaxation time.  Moreover, it often captures, in its 
simplicity, the basic ingredients of the underlying physics.  In this 
section, we will give a formulation of the finite-frequency 
fluctuational excess conductivity in terms of this formalism.  Since 
we are dealing with YBCO, anisotropy has to be taken into account.  
Due to the moderate anisotropy of this compound, we choose a 
description in terms of uniaxially anisotropic three-dimensional 
superconductor (we will come back to the consistency of this choice in 
the Discussion).\\
Calculation of fluctuational conductivity can be performed within 
several formalisms, such as the Kubo formula, the Boltzmann equation\cite{mishonov2,puica}
and the correlation-function formalism. We adopt here the model based on the 
correlation-function formalism,\cite{dorsey} extended to include the 
uniaxial anisotropy.\cite{neri} This approach is based on the 
following very general points:\\ 
{\em 1-} The GL functional is written as an expansion in $\psi$ (the 
order parameter).\\
{\em 2-} The response to a time-varying field ${\bf A}(t)$ (${\bf A}$ 
is the vector potential) is determined by the current operator 
averaged with respect to the noise (represented below by the 
brackets), and it can be expressed as a function of the correlation 
function of the order parameter $C\left( {\bf r},t;{\bf r}^{\prime 
},t^{\prime }\right) =\left\langle \psi \left( {\bf r},t\right) \;\psi 
^{\star}\left( {\bf r}^{\prime },t^{\prime }\right) \right\rangle $ at 
equal times:
\begin{equation}
\left\langle J_x\left( t\right) \right\rangle =- \frac{\hbar 
e^{\star}}{m_{x}}\int \frac{d^3 q}{\left( 2\pi \right) ^3}q_xC\left[ 
{\bf k}={\bf q}- \frac {e^{\star}} {\hbar} {\bf A}\left( t\right) 
;t,t\right] \
\label{corrente}
\end{equation}
where the momentum dependence has been shifted from ${\bf k}$ to the 
new vector ${\bf q}={\bf k}+\left( e^{\star}/\hbar \right) {\bf 
A}\left( t\right) $, $e^{\star}=2e$ is twice the electronic charge, 
and $m_{j}$ are the masses of the pair along the main crystallographic 
directions.\\
Standard calculations of the fluctuation conductivity are performed 
using the following conditions:\\
{\em i)} integration can be extended to the full 
momentum space, being the contribution at 
$q=0$ the most diverging close to $T_{c}$;\\
{\em ii)} terms containing powers of $\psi$ 
higher than 2 are dropped in the GL functional.\\
Using these hypotheses, one gets in a uniaxially anisotropic 3D 
superconductor, and in absence of a dc magnetic field, the in-plane 
excess conductivity (in Sist\`eme International units): 
$\Delta\sigma_{\infty}(\epsilon,\omega)=\frac{e^{2}}{32\hbar 
\xi_c(0)\epsilon^{1/2}} \;\left[ S_{+}\left( w\right) +{\rm i}\,S_{-}\left( 
w\right) \right]$, where $\epsilon = \ln(T/T_c)$ is the reduced 
temperature, \cite{gorkov} $\xi_c(0)$ is the out-of-plane 
zero-temperature GL coherence length and $w=\omega \tau$ with 
$\tau=\tau_{0}/\epsilon$ the temperature dependent GL relaxation time.  
$S_+ (w)$ and $S_- (w)$ are the scaling functions as can be found in 
Ref.\onlinecite{dorsey}, and they have the property that $S_{+} (w\to 
0)\simeq 1-\frac{w^{2}}{16}$ and $S_{-} (w\to 0)\simeq \frac{w}{6}$.  
We will refer to this result as ``infinite cutoff approximation'' (IC), and we 
will use the subscript ``$\infty$'' for it.  The above reported expression 
for $\Delta\sigma_{\infty}$, originally derived\cite{schmidt} for isotropic 
superconductors, was found to well describe microwave fluctuational 
conductivity in conventional superconductors.\cite{oldmicro} Moreover, 
a scaling property was found close to $T_{c}$ in YBCO films in 
swept-frequency measurements,\cite{booth} even if the temperature 
dependence of $\tau$ was markedly different from the GL prediction.\\
At temperatures sufficiently higher than $T_{c}$ the approximation
{\em i)} above should be somehow relaxed.  In fact, Eq.\ref{corrente}
should be integrated only over momenta that give a physical
contribution: modes with $q>\xi_{0}^{-1}$, where $\xi_{0}$ is the
temperature-independent coherence length, should be discarded.  This
fact was recognized early in the study of the dc excess
conductivity,\cite{johnson} and largely applied to various
superconductors.\cite{pavuna,silvaPhC,carballeira,silva,johnson,gauzzi}
However, the absence of explicit expressions for the complex excess
conductivity has limited the application of such approach to a few
experimental results, and numerical methods had to be 
used.\cite{nerisatt,nerimos} The usual way to discard
high-momentum modes is to introduce a spectral cutoff
$q_{j}^{max}=\Lambda_{j}\xi_{j0}^{-1}$ for each crystallographic direction
in Eq.\ref{corrente} and to calculate the cutoff excess
conductivity.  In order to simplify the resulting expression, and to
reduce the number of fitting parameters, we choose a single cutoff 
number imposing:
$\sqrt{\sum\limits_{j=1,..3}[q_j^{max}\xi_j(0)]^{2}}<\Lambda$
(Ref.\onlinecite{notaxi}).  The resulting complex excess conductivity
can be calculated explicitly,\cite{silvaEPJB} and for a uniaxial
superconductor the result reads:

\begin{eqnarray}
	\Delta\sigma_{3D}=\frac{e^{2}}{32\hbar \xi_c(0)\epsilon^{1/2}} 
	\;\frac{16}{3\pi w^{2}} \left\{atn\; K 
	-\left(1-{\rm i}w\right)^{3/2}\;atn\left(\frac{K}{\sqrt{1-{\rm i}w}}\right)+ 
	\right.  \nonumber \\
	\left.  +{\rm i}w\left[ \frac{K}{2\left(1+K^{2}\right)}-\frac{3}{2} atn 
	\; K \right] \right\}
\label{sigma3D}
\end{eqnarray}

where $K = \Lambda/\epsilon^{\frac{1}{2}}$.  In order to clarify some 
of the main features of Eq.\ref{sigma3D}, it is useful to introduce the small $w$
expansion. Up to terms of order $w^{2}$ one has:

\begin{eqnarray}
 \Delta\sigma_{1}(\epsilon,w<<1)\simeq \frac{e^{2}}{32\hbar 
 \xi_c(0)\epsilon^{1/2}} \left\{ 
 \frac{2}{\pi}\left[atnK-\frac{K}{\left(1+K^{2}\right)^{2}}\left(1+\frac{5}{3}K^{2}\right)\right]+ 
 o(w^{2})\right\}
 \label{sigma1approx} 
 \end{eqnarray}
 
 \begin{eqnarray}
 \Delta\sigma_{2}(\epsilon,w<<1)\simeq \frac{e^{2}}{32\hbar 
 \xi_c(0)\epsilon^{1/2}}  \frac{w}{6} \frac{2}{\pi} 
 \left[atnK+\frac{K}{\left(1+K^{2}\right)^{3}}\left(K^{4}-\frac{8}{3}K^{2}-1\right)\right]
 \label{sigma2approx} 
 \end{eqnarray}
 
 Like in the IC approximation, the leading term in the real part is frequency
 independent and the corresponding term in the imaginary part is
 proportional to the frequency.  On the other hand, the cutoff affects
 strongly the temperature dependences, in particular on the real part (see
 also the discussion in Ref.\onlinecite{silvaEPJB}). It is then expected 
 that the analysis of the data in the region $w<<1$ can be used to 
 easily discriminate between the IC and the finite-cutoff expressions.\\
Regarding the assumption {\em ii)}, it becomes less and less
acceptable approaching very closely $T_{c}$.  In order to take into
account the critical fluctuation regime, Dorsey\cite{dorsey} has
released that assumption by using the Hartree approximation for the
quartic term in $\psi$, and keeping the validity of {\em i)} above. 
The resulting expression of the complex excess conductivity remained
unaffected, including the scaling functions $S_+ (x)$ and $S_- (x)$,
and the effect of the extended approach was to change the temperature
dependence of the reduced temperature $\epsilon$ (renormalization)
according to:

\begin{equation}
\tilde{\epsilon}= \epsilon\; \frac{\epsilon}{\eta}\; \left[ 1+ \left( 
1+\frac{\epsilon} {\eta} \right)^{1/2} \right]^{-2}
\label{epstil} 
\end{equation}

When one takes into account the anisotropy,\cite{neri} in the 
preceeding expression $\eta=l^{2}\kappa^{4}\gamma^{2}\xi_{ab}^{2}(0)$, 
$l=\left( e^{2} \,\mu _{0} \,k_{B} T/4\pi \,\hbar ^{2} \right) $, 
$\mu_{0}$ is the magnetic permeability, $\gamma = \xi_{ab}(0) / 
\xi_{c}(0)$ is the anisotropy factor and $\kappa = 
\lambda_{ab}(0) / \xi_{ab}(0)$ is the Ginzburg parameter, and 
$\xi_{c}(0)$ and $\lambda_{ab}(0)$ are the zero-temperature values of the 
GL out of plane coherence length and in-plane penetration depth, 
respectively.  Summarizing, in the Hartree approximation the 
fluctuational excess conductivity has the same form as in the IC, unrenormalized 
limit, with the replacement $\epsilon\rightarrow \tilde{\epsilon}$.  
We remark that close to $T_{c}$ the temperature dependence of the GL 
relaxation time changes to $\tau\sim\epsilon^{-2}$, showing the 
so-called slowing down of fluctuations.\\
We note that the two above-described modifications to the IC, 
unrenormalized 
approximation have a significant relevance in different temperature 
ranges: the renormalization is relevant only close to $T_{c}$ 
($T-T_{c}\lesssim$ 2 K), while the influence of the cutoff in 
Eq.\ref{sigma3D} shows up above that temperature range.  We then cast 
together the two modifications by substituting $\epsilon$ with 
$\tilde{\epsilon}$ in Eq.s 
\ref{sigma3D},\ref{sigma1approx},\ref{sigma2approx}.\\
In concluding this section, we ought to mention that the cutoff 
approach is still somehow debated.  However, it is believed to be a 
fundamental aspect of the GL theory.\cite{landau,mishonov,werthamer} In the 
specific case of the dc, 2D excess conductivity, we have thoroughly 
discussed the similarities between the GL cutoffed model and the 
microscopic theory,\cite{reggiani} and we have shown that the 
GL expression with cutoff reproduces nearly exactly the microscopic theory for 
clean superconductors, when the appropriate cutoff number is chosen.\\
Finally, we comment on the fact that the momentum cutoff is not the 
only possible approach in order to extend the validity of the GL 
theory beyond the vicinity of $T_{c}$.  In fact, an alternative choice 
is the so-called ``total energy'' cutoff,\cite{carballeira} where it is 
the kinetic + localization energy of the (fluctuating) Cooper 
pair that is limited to some value.  The results for the ac excess 
conductivity are formally identical to the ones reported in Section 
\ref{th}, with the substitution 
$\Lambda\rightarrow\sqrt{c^{2}-\epsilon}$, with $c$ a new cutoff 
number, as noted earlier,\cite{silvaEPJB} in analogy with the other 
physical quantities.\cite{mosqueira} Eq.\ref{sigma3D} modified in 
order to use a total energy cutoff would exhibit a stronger decrease 
of the excess conductivity at high $\epsilon$, with vanishing 
$\Delta\sigma$ at $\epsilon=c^{2}$.

\section{Discussion}
\label{disc}

\begin{figure}
\centerline{\psfig{figure=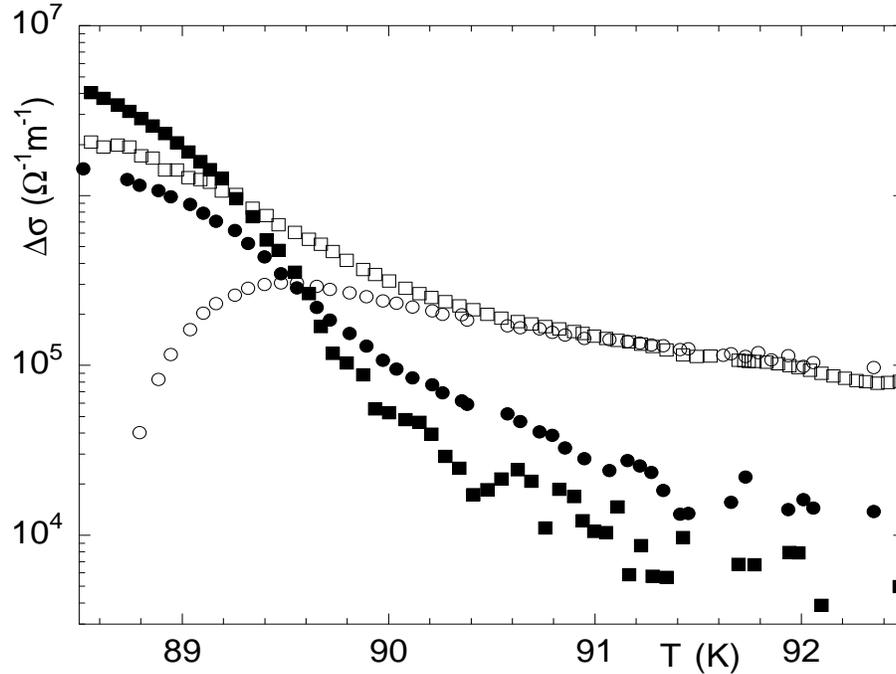,height=9cm,width=12cm,clip=,angle=0.}}
\caption{Complex excess conductivity as a function of the temperature 
in sample YA. Open symbols: $\Delta\sigma_{1}$, full symbols: 
$\Delta\sigma_{2}$.  Squares: 23.9 GHz, circles, 48.2 GHz.  Above 
$\sim $ 90.5 K, $\Delta\sigma_{1}$ does not exhibit 
significant frequency dependence.  It is also seen that 
$\Delta\sigma_{2}$ scales approximately as $\omega$.}
\label{fig2}
\end{figure}

\begin{figure}
\centerline{\psfig{figure=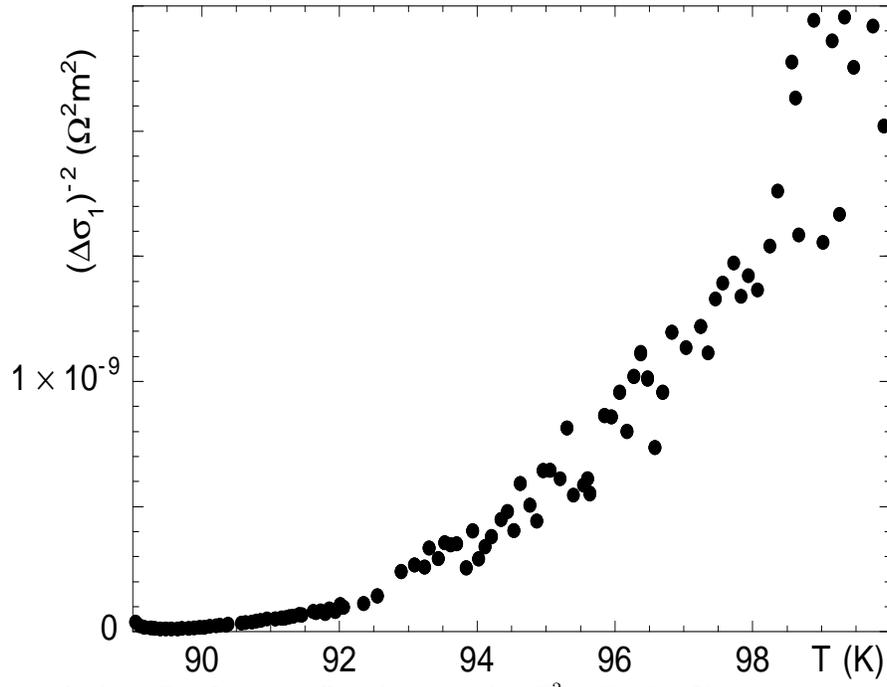,height=9cm,width=12cm,clip=,angle=0.}}
\caption{Comparison with plain 3D infinite cutoff predictions,  
$\left(\Delta\sigma_{1}\right)^{-2}\propto T-T_{c}$.  Clear curvature 
in $\left(\Delta\sigma_{1}\right)^{-2}$ is evident, in 
disagreement with the infinite cutoff approximation.  
Data refer to sample YA at 48.2 GHz.}
\label{fig3}
\end{figure}

\begin{figure}
\centerline{\psfig{figure=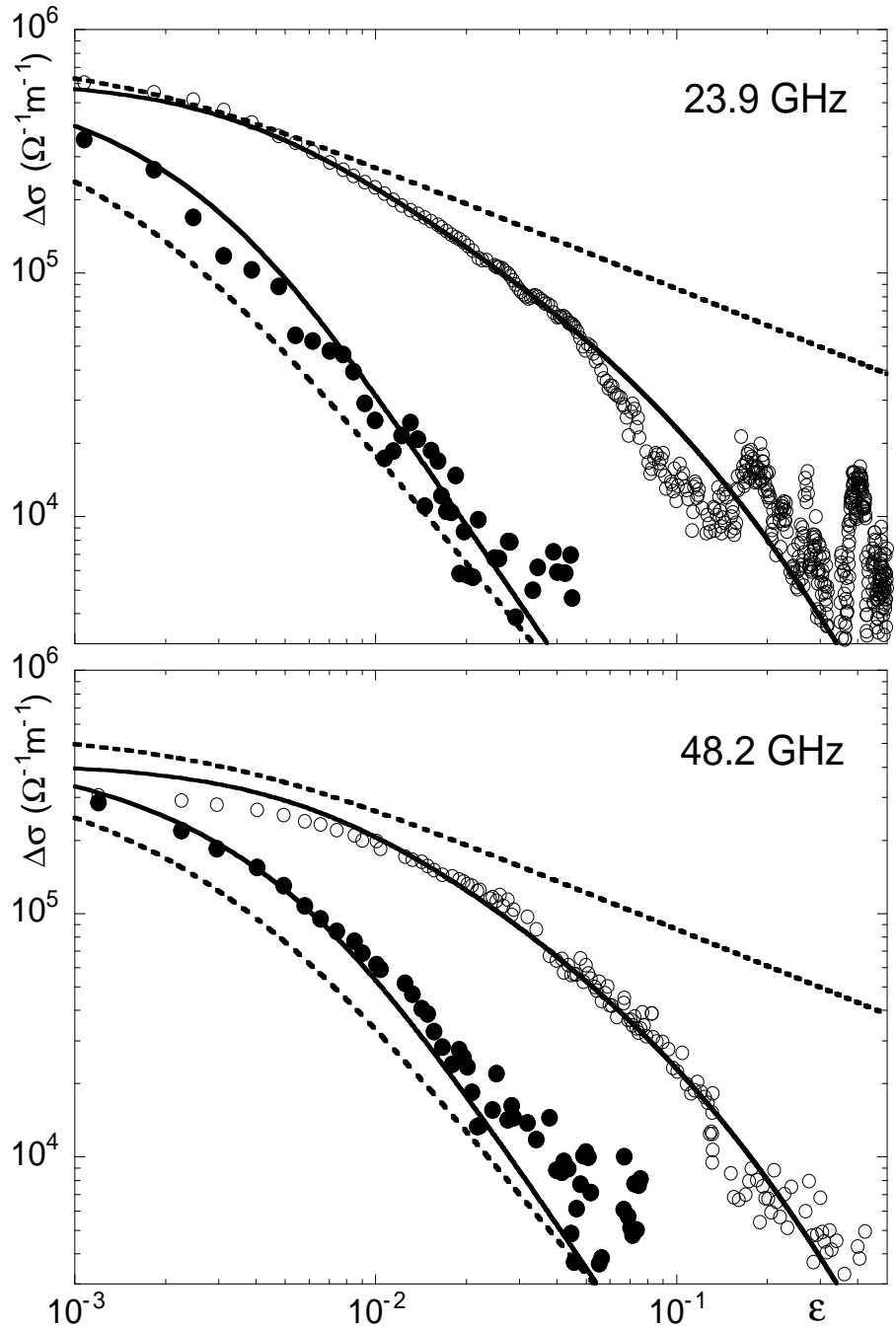,height=18cm,width=12cm,clip=,angle=0.}}
\caption{Complex excess conductivity as a function of the reduced 
temperature $\epsilon = \ln(T/T_c)$.  Open symbols: 
$\Delta\sigma_{1}$, full symbols: $\Delta\sigma_{2}$.  Upper panel: 
23.9 GHz, lower panel: 48.2 GHz.  Data taken on sample YA. Full lines 
are fits with Eq.\ref{sigma3D}, with parameters as in Table I. 
Dashed lines are unrenormalized, infinite cutoff, best fits, with $\tau_{0}$= 27 fs and the 
scale factor $\xi_{c}(0)$=2.8\AA  chosen to match the experimental 
$\Delta\sigma_{1}$ at 23.9 GHz.  The conventional model is clearly 
inadequate in order to simultaneously fit the four experimental curves.}
\label{fig4}
\end{figure}

\begin{figure}
\centerline{\psfig{figure=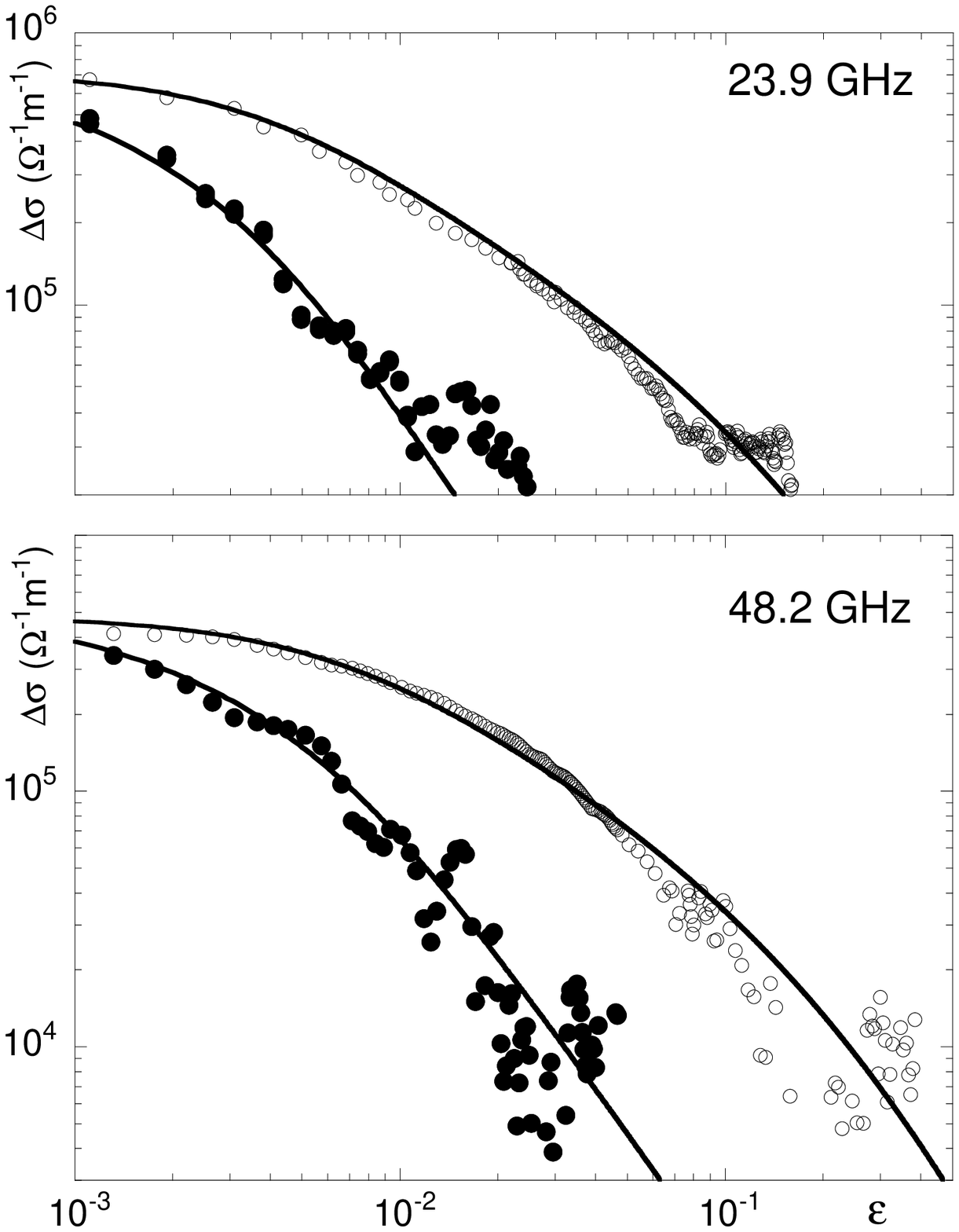,height=16cm,width=12cm,clip=,angle=0.}}
\caption{As in Figure \ref{fig4}, but data are taken on sample YB. Open 
symbols: $\Delta\sigma_{1}$, full symbols: $\Delta\sigma_{2}$.  Upper 
panel: 48.2 GHz, lower panel: 23.9 GHz.  Full 
lines are fits with Eq.\ref{sigma3D}, with parameters as in Table I. }
\label{fig5}
\end{figure}

We now discuss our data in light of the theoretical frame of 
Sec.\ref{th}.  We first identify the temperature region where $w<<1$ 
in our data.  In Fig.\ref{fig2} we present the complex excess 
conductivity at 23.9 and 48.2 GHz in sample YA as a function of $T$.  
As can be seen, at temperatures above 90.5 K the frequency dependence 
is nearly absent in $\Delta\sigma_{1}$, indicating that the GL 
relaxation rate $1/\tau_{0}$ is much larger than our measuring 
frequency (see Eq.s \ref{sigma1approx}).  Consistently (see Eq.s 
\ref{sigma2approx}), the imaginary part $\Delta\sigma_{2}$ scales 
approximately as the ratio of the measuring frequencies.  These 
features represent the experimental indication that, for most of the 
temperature range here explored, the small frequency regime is 
reached.\\
We now show that the simple IC model is not compatible with our 
data.  Using the fact that, within $1\%$, one has $\epsilon\simeq 
\frac{T}{T_{c}}-1$
up to $T=1.1T_{c}$, the IC approximation in the small $w$ regime 
predicts $\Delta\sigma_{1,\infty}^{-2}\propto T-T_{c}$ in a rather wide 
temperature range, as opposed to the cutoff expression (see 
Eq.\ref{sigma1approx}), that removes this simple temperature 
dependence.\\
In Fig.\ref{fig3} we plot the data as 
$\left(\Delta\sigma_{1}\right)^{-2}$ vs.  $T$.  It is clear that our 
data do not follow a linear law in a significant temperature range, as 
would be expected from the IC approximation. The 
temperature and
frequency dependence of $\Delta\sigma$ further indicate (see later and 
Fig.\ref{fig4}) that the simple IC prediction does not 
describe our data. We conclude that a 
proper description for our data of the fluctuational conductivity must 
be found beyond the IC approximation.\\
We adopt the extension of the TDGL theory described in Sec.\ref{th}.  
In order to take into account temperatures far from $T_{c}$ as well as 
the approach to the critical region, we use in the full expression, 
Eq.\ref{sigma3D}, the renormalized reduced temperature, 
Eq.\ref{epstil}.  The resulting explicit expression for the complex 
$\Delta\sigma$ contains as parameters the cutoff number $\Lambda$ 
(effective mostly at high temperatures), the out-of-plane, 
zero-temperature GL coherence length $\xi_{c}(0)$ (which acts mainly 
as a scale factor), the GL relaxation time $\tau_{0}$ (determined 
mainly through $\Delta\sigma_{2}$), the number $\eta$ (which 
determines the reduced temperature range of the renormalized regime), 
or alternatively the zero-temperature, in-plane penetration depth 
$\lambda_{ab}(0)$, and $T_{c}$.\\
An extremely precise determination of $T_{c}$ is not essential, since 
the main point of this paper is the ``high'' temperature regime.  
Nevertheless, scaling theories\cite{dorsey} in a 3D superconductor 
suggest the following path for its evaluation: in fact, one has 
$\Delta\sigma_{1}(T\rightarrow T_{c}) =\Delta\sigma_{2}(T\rightarrow 
T_{c})$, so that the crossing point of the real and imaginary parts 
should give the critical temperature.  Unfortunately this 
determination can be easily affected by experimental errors in the 
evaluation of the absolute value of the excess conductivity such as 
those mentioned in Sec.\ref{exp}.  We have then chosen to keep $T_{c}$ 
as a parameter of the fit, keeping in mind that it cannot be 
noticeably different from the cross-point $T^{\times}$ of the real and 
imaginary parts of $\Delta\sigma$.  The values of the crossing 
temperature $T^{\times}$ and the fitted $T_{c}$ are reported in the 
Table, and good agreement is found.\\
All the parameters are fixed by the simultaneous fitting of four 
experimental curves in each sample: real and imaginary parts at two 
different measuring frequencies.\\
Fits are presented in Fig.s \ref{fig4} and \ref{fig5} as continuous 
lines.  It is clear that the extended model well describes our data 
for both microwave frequencies, as opposed to the failure of the 
IC approximation (dashed lines in Fig.\ref{fig4}).  In particular, it 
is the temperature dependence at large $\epsilon$ that clarifies the 
role of the cutoff, while the slowing down of fluctuations at 
$T_{c}$ is described by the renormalization.\\
The so-obtained fitting parameters are summarized in Table I. The 
zero-temperature c-axis coherence lengths, $\xi_{c}(0)$, are in 
agreement with commonly accepted values.\cite{notaprop} The renormalization 
parameter $\eta\simeq 7\cdot10^{^-3}$ (corresponding to a GL 
$\lambda_{ab}(0)\simeq $900 \AA) indicates that in a region $\sim$0.6 K wide close to 
$T_{c}$ the corrections to the $\psi^{4}$ cannot be neglected.  The 
cutoff number $\Lambda$ is of the correct order of magnitude, and 
appears to be sample-dependent.  We already found\cite{silva} (from dc 
conductivity) that in slightly overdoped 
Bi$_{2}$Sr$_{2}$CaCu$_{2}$O$_{8+x}$ crystals there was a 
doping-dependence of the cutoff number.  It would be interesting to 
look for a similar dependence in YBCO. We note that the obtained parameters 
are consistent with the 3D treatment here employed: in fact, in order 
to neglect the layered structure one needs the anisotropy parameter
$\left(\frac{2\xi_{c}}{t}\right)^{2}$ ($t\simeq$ 7 \AA  is the spacing between 
superconducting planes in YBCO) to be sufficiently larger than 
$\Lambda^{2}$ (Ref.s \cite{mishonov,puica}). With our parameters we 
obtain a factor $\sim$ 3 between the two quantities, that points to 
an essentially 3D nature of YBCO.\\
The GL relaxation time is found to be slightly sample dependent, in 
agreement with results obtained from the analysis at 
microwave\cite{booth} and radio\cite{nakielski} frequencies and from 
combined excess diamagnetism and conductivity 
measurements.\cite{ramallo} The numerical value for $\tau_{0}$ is 
noticeably larger than the BCS estimate, $\tau_{0} =\frac{\hbar \pi 
}{16 k_{B} T}\simeq$ 17 fs (having taken $T=T_{c}$), in qualitative 
agreement with Ref.\onlinecite{booth}.  A possible interpretation of 
this finding comes from the unconventional nature of the pairing in 
cuprates.  In fact, it is expected\cite{mishonov} that in a d-wave 
superconductor the GL relaxation time is larger than the BCS, s-wave 
estimate.  It is clear that this intriguing point deserves further 
experimental work in the future, possibly expanding the frequency 
range explored.\\

\section{Conclusions}
\label{conc}
We have measured the excess complex conductivity at two microwave 
frequencies in very high quality YBCO films.  The temperature 
dependence of the data in the region $\omega\tau<<1$ directly shows 
that the established results of the time-dependent Ginzburg Landau 
theory within the IC approximation do not describe our data.  We 
have incorporated in the TDGL theory the renormalization (R) of the 
reduced temperature and the short-wavelength cutoff (C), obtaining a 
single explicit expression for the finite-frequency excess 
conductivity.  With these modifications, the RC-GL theory describes in 
full our data, in the temperature and in the frequency dependence.  
With a reduced parameter set we are able to fit at once four 
experimental curves for each sample.  The very good fitting and 
remarkable consistency of the parameters support the idea that the 
proposed modifications of the TDGL model capture the essential 
features of the complex excess conductivity in YBCO. We have estimated 
the out-of-plane coherence length, the in-plane penetration depth and 
the GL relaxation time.  The latter appears longer than the BCS s-wave 
value.  This point, in qualitative agreement with the expectations for 
d-wave superconductors, deserves further experimental work in the 
future.\\

{\em $-$ Note added. After completing this manuscript, a related work 
appeared, discussing a possible effect of the cutoff close to $T_{c}$ 
within a model for the ac fluctuation conductivity similar to the one 
here presented.\cite{peligrad} Such effect should manifest itself in 
an increasing T$^{\times}$ with the frequency. This is qualitatively 
consistent with our data, see the Table.}

\section{acknowledgements}
We thank D.Neri for help and for many useful discussions in the early 
stage of this work, M.W.Coffey and T.Mishonov for useful discussions 
and stimulating correspondence, R.Raimondi for useful discussions.  
Work in Salerno was partially supported by a MURST-COFIN 2000 
project.\\

\begin{table}
\begin{tabular}{ccc}
 & Sample YA & Sample YB \\ 
$T^{\times}_{23.9}$(K) & 89.3 & 89.25\\
$T^{\times}_{48.2}$(K) & 89.5 & 89.3\\
${T}_c$(K) & 89.45 & 89.25\\ 
$\xi_{c}(0)$(\AA) &  3.3& 2.8\\ 
$\tau_{0}$(fs)& 27 & 29\\ 
$\eta$(\AA) &  7 10$^{-3}$ & 7 10$^{-3}$\\ 
$\Lambda$&  0.55 & 0.65
\end{tabular}
\caption{Crossing temperatures of real and imaginary parts 
$T^{\times}$ at the two measuring frequencies (see text) and fit 
parameters: critical temperatures $T_{c}$, zero-temperature c-axis 
coherence length $\xi_{c}(0)$, GL relaxation time $\tau_{0}$, 
renormalization parameter $\eta$ (see Eq.\ref{epstil}), and cutoff 
number $\Lambda$.}
\label{t1}
\end{table}

% %----------------- FIGURE 1 --------------------
\newpage
% %----------------- FIGURE 2 --------------------
%\newpage
% 
% %----------------- FIGURE 3 --------------------
%\newpage
% 
% %----------------- FIGURE 4--------------------
%\newpage
% 
% %----------------- FIGURE 5 --------------------
%\newpage
% 


\begin{references}
\bibitem{notaxi} For the sake of compactness, throughout the paper we 
use only the GL zero-temperature coherence length, $\xi(0)$, instead 
of the microscopic coherence length $\xi_{0}=\xi(0)/0.74$.  As a 
consequence, the cutoff $\Lambda$=0.74 means a momentum cutoff 
$Q=\xi_{0}^{-1}$. We also note that our choice of the cutoff 
would correspond to a kinetic energy cutoff 
$E_{max}<\Lambda^{2}\frac{\hbar^{2}}{2m^{*}\xi^{2}(0)}$
\bibitem{notaprop} We recall that $\xi_{c0}$ acts as a scale factor.  
All the errors in the proportionality factors in the conversion between 
the quality factor and the resonance shift to the complex conductivity 
are absorbed into $\xi_{c0}$
\end{references}
\end{document}